\documentclass[runningheads,envcountsame,a4paper]{llncs}
\usepackage{graphicx}

\usepackage{tikz}
\usepackage{amsmath}
\usepackage[most]{tcolorbox}
\usepackage{comment}
\usepackage{epsfig,endnotes}
\usepackage{url}
\usepackage[numbers]{natbib}
\usepackage{multirow}
\usepackage{graphics}
\usepackage{cite}
\usepackage{xcolor}
\usepackage[para]{footmisc}
\usepackage{lipsum}
\usepackage{tabularx,booktabs}
\usepackage[section]{placeins}
\usepackage{graphicx}
\usepackage{comment}
\usepackage{enumitem}
\usepackage{float}
\usepackage[misc]{ifsym}
\restylefloat{table}

\usepackage{caption}
\usepackage{subcaption}
\usepackage{mwe}    % for example figures!
\renewcommand\thefigure{\arabic{figure}}
\DeclareCaptionLabelFormat{simplab}{Figure~#2}
\DeclareCaptionFormat{nocap}{}
\DeclareCaptionSubType*{figure}

\usepackage{array,multirow,booktabs,mathtools,tabulary,xcolor}

\usepackage{colortbl}
\usepackage[T1]{fontenc}
\usepackage[utf8]{inputenc}
\let\belowcaptionskip\abovecaptionskip

\arrayrulecolor{blue!70} 

\makeatletter
 \def\@textbottom{\vskip \z@ \@plus 1pt}
 \let\@texttop\relax
\makeatother

\usepackage{filecontents}
\begin{document}

\title{Are Current CCPA Compliant Banners Conveying User's Desired Opt-Out Decisions? An Empirical Study of Cookie Consent Banners}

\author{Torsha Mazumdar\textsuperscript{*}, Daniel Timko\thanks{Both authors contributed equally.}, Muhammad Lutfor Rahman  }

\institute{California State University San Marcos \\
\email{mazum001@csusm.edu, timko002@csusm.edu, mlrahman@csusm.edu}}

\toctitle{Are Current CCPA Compliant Banners Conveying User's Desired Opt-Out Decisions? An Empirical Study of Cookie Consent Banners}
\tocauthor{Torsha Mazumdar, Daniel Timko, Md Lutfor Rahman}

\maketitle

\begin{abstract}
The California Consumer Privacy Act (CCPA) secures the right to Opt-Out for consumers in California. However, websites may implement complex consent mechanisms that potentially do not capture the user’s true choices. We investigated the user choices in Cookie Consent Banner of US residents, the plurality of whom were from California, through an online experiment of 257 participants and compared the results with how they perceived to these Cookie Consent Banner. Our results show a contradiction between how often participants self-report their Opt-Out rates and their actual Opt-Out rate when interacting with a complex, CCPA-compliant website. This discrepancy expands the context with which modern websites may implement the CCPA without providing users sufficient information or instruction on how to successfully Opt-Out. We further elaborate on how US residents respond to and perceive the GDPR-like Opt-In model. Our results indicate that even though very few consumers actually exercised their right to Opt-Out, the majority of US consumers desire more transparent privacy policies that the current implementation of CCPA on websites lacks.

\end{abstract}

\section{Introduction}

Over the past decade, there has been a significant increase in global awareness regarding data privacy. Personal information, as well as data on preferences, interests, and browsing behavior, is being captured, collected, sold to other companies, and analyzed in order to deliver personalized user experiences. While consumers are often curious or irritated by the sudden appearance of unsolicited advertisements on their screens or the constant influx of promotional emails in their inboxes, businesses also face the daunting challenge of ensuring data protection.

In response to escalating concerns about data privacy, the California Consumer Privacy Act (CCPA), which came into effect on January 1, 2020, grants consumers greater control over the personal information collected by businesses. The CCPA establishes regulations that offer guidance on the implementation of the law. Similar in many aspects to the European Union's General Data Protection Regulation (GDPR), the CCPA and GDPR share a common goal of safeguarding personal data by ensuring its fair, lawful, and transparent use.

However, despite these regulations, numerous businesses either fail to comply or intentionally create convoluted opt-out processes that are challenging for users to navigate~\citep{liu2022opted}. Moreover, many users either remain unaware of their rights, exhibit reluctance to read privacy policies thoroughly, or simply do not prioritize data privacy to the same extent.

In this study, we conducted an online experiment involving 257 US residents to investigate how users react to consent notices. To enhance the experiment's realism, we employed a minor deception technique in the online task. Instead of explicitly focusing on the cookie banner consent popup, we instructed users to consider the browser's security indicators. By employing this minor deception, we aimed to minimize priming effects and obtain results that truly reflect users' authentic online behavior~\citep{salah2012deceive}. Additionally, to complement the online experiment, we administered a survey to gather users' perspectives on their data privacy rights and the data practices of corporations.

We further analyze the impact of various implementation choices commonly employed on websites subject to the CCPA on users' ability to make informed consent decisions~\citep{DBLP:journals/corr/abs-2009-07884}. It is important to note that the CCPA specifically applies to California residents and businesses operating within the state. Unlike the Opt-In model adopted by its European Union counterpart GDPR, the CCPA includes specific regulations regarding the right to Opt-Out. Through our online experiment, we closely observe and assess users' responses to the Opt-In and Opt-Out options, as well as capture their thoughts and opinions regarding these choices. In summary, we have the following contributions to our study. 
\begin{enumerate}
\item We design a website that uses the pretense of a security indicator setup task to record the activity of participants with randomly assigned cookie consent mechanism. We conducted a user study to determine how different consent mechanisms affect user's Opt-Out rates.

\item 

We compare the rates of actual Opt-Out against the perception of how often the average user states they want to Opt-Out. 
\item 

We analyze the user perspectives on the Opt-In model and their awareness of CCPA regulations among US residents.

\end{enumerate}

\section{Related Works}

\textbf{CCPA and consent mechanisms} 
The CCPA introduces a crucial right known as the \textit{Right to Opt-Out of Sale or Share}, which is commonly implemented by offering users the option to withdraw their consent by clicking on a button or link within a website or application. It grants California residents the authority to file complaints with the Office of the Attorney General (OAG) in cases of suspected CCPA violations. However, it can be challenging for an average individual to ascertain whether a specific website is subject to or exempt from the CCPA~\citep{Nortwick}.

O'Connor et al.~\citep{DBLP:journals/corr/abs-2009-07884} conducted a study that revealed the significant impact even minor differences in implementation decisions can have on user interactions with consent notices that appear on screens. The CCPA, which draws inspiration from the GDPR, mandates that Opt-Out links for sale must be \textit{clear and conspicuous}. However, it has been observed that many websites adopt design and implementation choices that appear to \textit{negatively} impact user privacy. Work by Utz et al.\citep{Utz2019} explored how the developers use placement of notifications, notification choices, and persuasive design techniques like dark patterns and nudging to influence consent decisions. Dark patterns are malicious design choices that direct users into certain behavior or choices. These design implementations can be used to reduce Opt-Out rates by providing a link to the Opt-Out mechanism on a separate web page or requiring users to scroll to the bottom of the page to find the link, rather than offering a direct link.

Additionally, nudging techniques, low-contrast font colors, and smaller font sizes are used to divert users' attention. Some websites even place the Opt-Out mechanism solely within their privacy policy, disregarding the CCPA guideline that specifies the need for an Opt-Out link on the homepage of the website.

In their study, Chen et al.~\citep{FightingtheFog} conducted an analysis of 95 privacy policies from popular websites and discovered inconsistencies among them. They observed that not all disclosures provided the level of clarity mandated by the CCPA when describing data practices. Moreover, their findings indicated that different wording choices influenced how consumers interpreted the data practices of businesses and their own privacy rights. The presence of \textit{Vagueness and ambiguity} in privacy policies significantly hampers consumers' ability to make informed choices~\citep{FightingtheFog}.

It is worth noting that even for thoroughly tested designs, consumer education remains crucial for several reasons. Firstly, it helps \textit{raise awareness among users, communicates the purpose of privacy icons, and dispels any misconceptions}~\citep{10.1145/3411764.3445387}. Additionally, unifying privacy choices in a single, standardized location would likely enhance user accessibility to these controls~\citep{10.1145/3313831.3376511}. The CCPA not only requires companies to provide privacy choices but also emphasizes the need to make these choices usable and user-friendly.

Opt-In systems require businesses to obtain an individual's \textit{express, affirmative, and informed} consent before sharing or processing their data. In contrast, the Opt-Out rule places the responsibility on the individual to safeguard their own data~\citep{Park2020TheCW}. Participants' responses can be influenced by the way questions are framed. Merely presenting the question as an Opt-Out instead of an Opt-In, or vice versa, often leads to different privacy preferences being recorded. Consent choices are frequently displayed with a pre-selected "Yes" or "Accept" response, exploiting individuals' inattention, cognitive biases, and tendency for laziness~\citep{Gerald}.

\section{Methodology}
Our study methodology was cleared by our university Institutional Review Board (IRB) and consisted of a presurvey questionnaire, online experiment and an exit survey. The average time required for this survey was 27 minutes. The time elapsed during the study was calculated from the start time of the presurvey questionnaire to the end time of the exit survey for each participant.

\subsection{Participant Recruitment}

Participants were recruited from the general US population. To be eligible, individuals had to be at least 18 years old and primarily English-speaking. While the main focus of this study is on the CCPA, which primarily affects the population of California, we also examined differences between residents in California and those in other states. We recruited participants for the "Study on Internet Users' Choice of Browser Security Indicators" through various channels, including flyers posted around the university campus, a call for participants on Craigslist, Reddit, and social media platforms such as Facebook, Instagram, and LinkedIn, as well as through the university email lists. As an incentive for participating in the study, two randomly selected participants out of the total 257 received a \$50 Amazon gift card.

\subsection{Presurvey Questionnaire}
The presurvey questionnaire comprised a consent form and a Qualtrics questionnaire that collected participants' demographic information. The informed consent form, serving as the initial page of the Qualtrics questionnaire, provided a detailed overview of the study. Only participants who explicitly consented proceeded to the subsequent steps of the study. Each participant was assigned a unique 4-digit random number, generated within Qualtrics, which was referred to as the Participant ID throughout the paper.

The demographic questionnaire consisted of multiple-choice questions aimed at capturing participants' age, gender, education, income, occupation, weekly internet usage hours, state of residence in the US, as well as their preferred devices, browsers, and operating systems.

The inclusion of experiment variables, such as age, gender, education, income, occupation, and allowed us to assess the representativeness of our sample population and the generalizability of our results. Examining the participants' internet usage shed light on their familiarity with navigating various websites, thereby contributing to the existing literature on the relationship between online literacy, privacy awareness, and willingness to share data. Additionally, the inclusion of experiment variables like devices, browsers, and operating systems enabled us to analyze the impact of these choices on the implementation of the consent notice and subsequently the users' responses. Given that the CCPA exclusively applies to California residents, the participants' state of residence within the US served as an important variable in our study, allowing us to compare the tendencies of the US population residing outside of California.

\subsection{Online Experiment}

Participants who completed the presurvey questionnaire were subsequently directed to our experiment website. In order to maintain the integrity of the study, we employed a method of deception by using advertisements and instructions on the last page of the presurvey questionnaire. Although the study aimed to examine participants' responses to consent notices, participants were led to believe that the study focused on their preferences for browser security indicators. The use of deception in research has been the subject of ethical debates, but it can be implemented safely when carefully framed and reviewed~\citep{salah2012deceive,10.1145/2470654.2466246,LAZAR201725}. Notably, several influential studies~\citep{Schechter07} have utilized deception to enhance the realism of experiments. The advantage of employing this deception is that it elicits more authentic responses from participants. If participants had been fully informed about the true purpose of our study, they might have approached the consent notices with increased attention and made choices that differ from their regular browsing tendencies. By incorporating an unrelated primary task in our experiment, we were able to observe how participants make consent decisions while simultaneously engaging in other prioritized tasks. This simulation reflects real-world scenarios, as users typically browse websites for personal, professional, or entertainment purposes rather than solely for accepting or rejecting cookies. The IRB granted approval for the use of deception in our study after determining that this aspect of the experiment would not cause any actual harm to participants. Furthermore, prior knowledge of the true purpose of the study was deemed likely to influence participants' behavior and potentially undermine the study's outcomes. At the conclusion of the exit survey, participants were provided with a debriefing that included information about the true purpose of the study, details about the CCPA, and suggestions for safeguarding their data.

\begin{figure}[htp]
    \centering
    \includegraphics[width=0.99\textwidth]{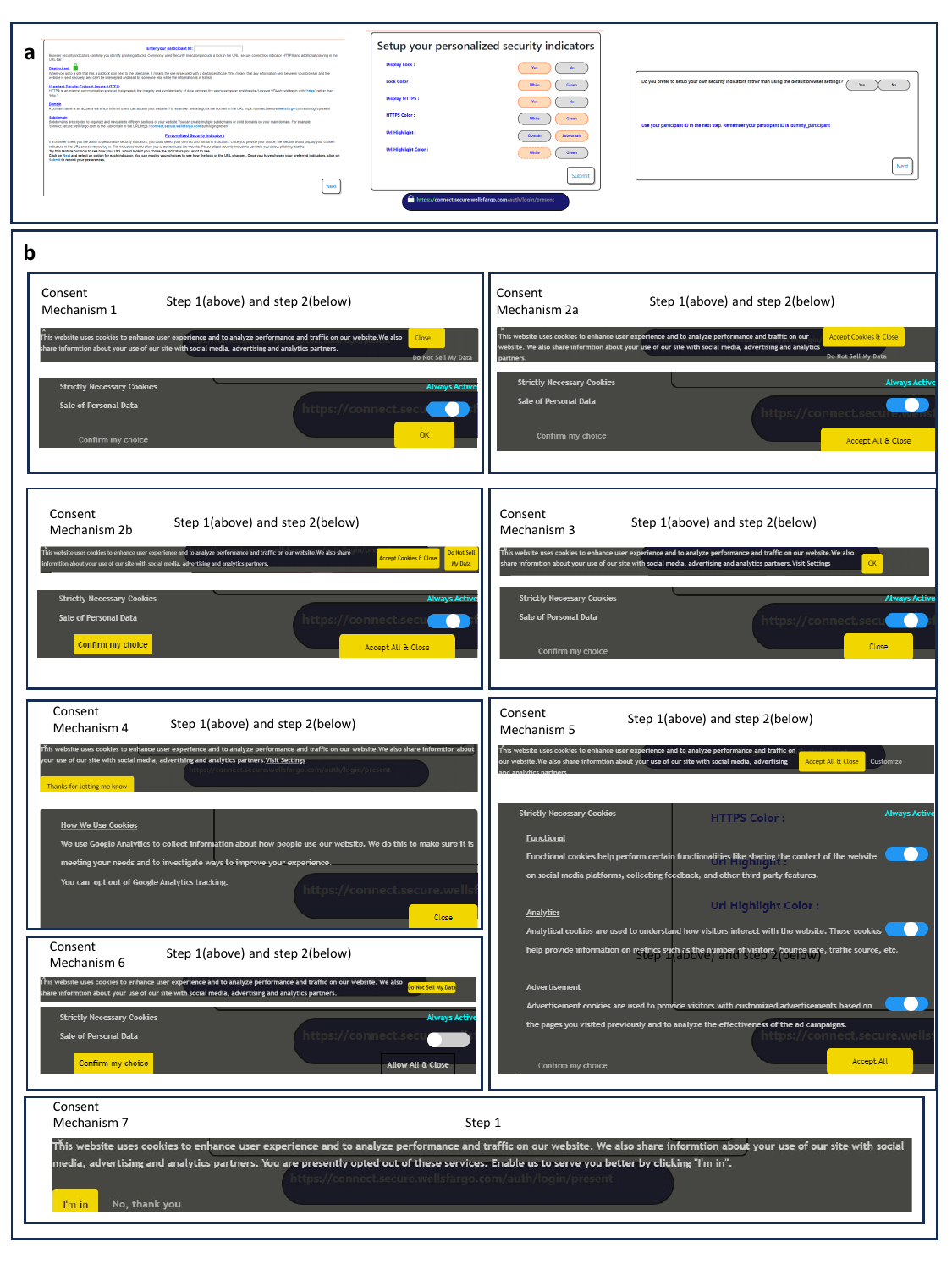}
    \vspace{-4mm}
    \caption{a) Left: Welcome page screen. Middle: Security indicator selection task screen. Right: Exit page screen where participants are directed to the exit survey. b) 7 cookie consent mechanism banners and their steps used in the online experiment.}
    \label{fig:primacytask}
\end{figure}

\vspace{-4mm}
\subsubsection{Primary Task for the Participants}

Participants were invited to participate in an online experiment framed as a study focusing on their preferred choices for browser security indicators. Browser security indicators serve the purpose of either alerting users to potentially suspicious URLs or assuring them of a secure connection. On the initial page of our website, participants were prompted to enter their Participant ID (generated in Qualtrics) and provided with context regarding the use of security indicators (see Figure~\ref{fig:primacytask}). They were then asked to indicate their preference for either the default browser security indicators or personalized ones (Figure~\ref{fig:primacytask}). Additionally, participants were asked to specify the type and color of the security indicators they would prefer to see in URLs (Figure~\ref{fig:primacytask}). The responses provided by participants were recorded and stored in our database. Although the users' choices were not directly related to our primary study on consent, they offer valuable insights into user preferences.

\vspace{-4mm}
\subsubsection{Consent Mechanisms}

According to the CCPA, it is required that the Opt-Out links on websites should be \textit{"clear and conspicuous"} on the homepage. As stated in the CCPA law document~\citep{ccpadocument}, \emph{"You may request that businesses stop selling or sharing your personal information (“opt-out”), including via a user-enabled global privacy control."}. However, many businesses have deliberately implemented mechanisms that impede users' ability to Opt-Out, resulting in lower Opt-Out rates. O'Connor et al.~\citep{DBLP:journals/corr/abs-2009-07884} conducted a study on the top 500 US websites and discovered significant deviations from the CCPA guidelines regarding Opt-Out mechanisms, deeming these implementations as \textit{"(Un)clear and (In)conspicuous."} They identified eight distinct types of Opt-Out controls used in popular websites, from which we selected and implemented five designs for our study. Additionally, some of our designs drew inspiration from popular customizable CCPA compliant banners~\citep{ccpacompliance}, which involve multiple decision steps for complete Opt-Out and offer toggle settings for opting out of specific content. In our experiment, we employed consent mechanisms that encompassed both the sharing and selling of personal information. The complete set of consent mechanism choices and their interpretations can be found in Table~\ref{tab:tab-b}.

\begin{enumerate}
\item \textbf{Consent Mechanism 1} - \textbf{"Do Not Sell My Data"} button and a \textbf{"Close"} button. Clicking on "Do Not Sell My Data" button would trigger a second consent banner to make specific choice ("Allow" or "Do Not Allow") in a toggle switch for sale of data and \textbf{"Confirm my choice"} or \textbf{"Close"}
\item \textbf{Consent Mechanism 2a} - \textbf{"Accept Cookies \& Close" }and a \textbf{"Do not sell my data"} button. Clicking on "Do Not Sell My Data" button would trigger a second consent banner to make specific choice ("Allow" or "Do Not Allow") in a toggle switch for sale of data and \textbf{"Confirm my choice"} or \textbf{"Allow All \& Close"}
\item \textbf{Consent Mechanism 3} - \textbf{"OK"} button to close and an in-line \textbf{"Visit Settings"} link. Clicking on "Visit Settings" link would trigger a second consent banner to make specific choice ("Allow" or "Do Not Allow") in a toggle switch for sale of data and \textbf{"Confirm my choice"} or \textbf{"Close"}
\item \textbf{Consent Mechanism 4} - \textbf{“Thanks for letting me know”} button and an in-line \textbf{“How We Use Cookies”} link. Clicking on "How We Use Cookies"  link would trigger a second consent banner with a "\textbf{Close"} button or an in-line \textbf{"opt out of Google Analytics tracking"} link.
\item \textbf{Consent Mechanism 5} - \textbf{"Accept All \& Close"} and a \textbf{"Customize"} button. Clicking on "Customize" button would trigger a second consent banner to make specific choice ("Allow" or "Do Not Allow") in a toggle switch for each kind of cookie(Functional, Analytics and Advertisement) and \textbf{"Confirm my choice"} or \textbf{"Accept All"}
\end{enumerate}

Although the aforementioned five consent mechanisms technically provide participants with the same options to either Opt-Out or remain opted-in by default for the sale or sharing of data, our study aims to investigate how the wording and implementation of each mechanism influence users' choices. Existing literature suggests that the format of questions plays a significant role~\citep{Johnson}, and our study seeks to verify this assertion. The remaining three Opt-Out controls described in O'Connor et al.~\citep{DBLP:journals/corr/abs-2009-07884} are not within the scope of this paper as they involve contacting third parties via email, filling out Opt-Out request forms, or adjusting account settings, which cannot be tested within a single experiment.

In their research, O'Connor et al.~\citep{DBLP:journals/corr/abs-2009-07884} also noted that nudging was commonly employed in direct links to subtly guide users away from successfully opting out. Digital Nudging~\citep{Weinmann2016} refers to a design approach where businesses highlight certain buttons to indirectly suggest, influence, or manipulate user behavior in a manner that benefits the businesses. In all of the above five consent mechanisms, we have incorporated nudging by highlighting options that do not allow users to Opt-Out. However, to examine all available conditions, we further expand Consent Mechanism 2 to include the following variation without nudging:

\textbf{Consent Mechanism 2b - Neutral:} Both the Opt-Out and accept options have the same design. Some websites employ an Anti-nudging design~\citep{DBLP:journals/corr/abs-2009-07884}, wherein only the Opt-Out option is presented without any further steps. Clicking on this button directly opts the user out. We implemented this design as the initial decision in \textbf{Consent Mechanism 6}.

While the CCPA mandates the right to Opt-Out, we also included an Opt-In mechanism based on the GDPR model, allowing us to compare and contrast the two types of mechanisms. In \textbf{Consent Mechanism 7}, a banner was displayed with an "I'm in" and a "No, thank you!" button, enabling participants to explicitly provide single step consent to the sale of their data.

Participants were directed to our experiment website through a link provided at the end of the presurvey questionnaire. The website was designed to display a consent banner on the screen after 3 seconds of a participant visiting the web page hosting the primary task. Following a completely randomized experiment design~\citep{LAZAR201725}, participants were randomly assigned one of the eight consent mechanisms developed for the study by selecting a random integer between 1 and 8. The assignment of consent mechanisms was counterbalanced \citep{ref1} by selecting a different random consent mechanism from the last recorded one, ensuring that the same random number was not repeatedly selected. This approach allowed us to gather sufficient data for analyzing participant behavior across each consent mechanism. It is important to note that we did not actually place any cookies on participants' devices or require participants to consent to the sale of their personal data. Instead, we used HTML forms to replicate commercial consent banners and recorded participants' responses in our experiment database.

For the readers' reference, snapshots of our website screens, showcasing the unique design of each banner, are included in Figure~\ref{fig:primacytask}.

\vspace{-4mm}
\begin{table*}[ht]
   \caption{Consent combinations and their interpretations. We present the Opt-Out path for each mechanism in bold.} 
   \label{tab:tab-b}
   \scriptsize
   \centering
\begin{tabular}{llcccc}
\hline
Default & \#CM & 1st Decision & 2nd Decision & \begin{tabular}[c]{@{}c@{}}2nd Decision \\ Toggle\end{tabular} & Interpretation \\ \hline
\multirow{5}{*}{Opt-In} & 1 & Close & NA & NA & Opt-In \\
 & 1 & Do Not Sell My Data & Confirm my choice & Always Active(On) & Opt-In \\
 & \textbf{1} & \textbf{Do Not Sell My Data} & \textbf{Confirm my choice} & \textbf{Do Not Allow} & \textbf{Opt-Out} \\
 & 1 & Do Not Sell My Data & OK & Allow & Opt-In \\
 & 1 & Do Not Sell My Data & OK & Do Not Allow & Opt-In \\ \hline
\multirow{5}{*}{Opt-In} & 2a 2b & Accept Cookies \& Close & NA & NA & Opt-In \\
 & 2a, 2b & Do Not Sell My Data & Allow All \& Close & Allow & Opt-In \\
 & 2a, 2b & Do Not Sell My Data & Allow All \& Close & Do Not Allow & Opt-In \\
 & 2a, 2b & Do Not Sell My Data & Confirm my choice & Allow & Opt-In \\
 & \textbf{2a, 2b} & \textbf{Do Not Sell My Data} & \textbf{Confirm my choice} & \textbf{Do Not Allow} & \textbf{Opt-Out} \\ \hline
\multirow{5}{*}{Opt-In} & 3 & OK & NA & NA & Opt-In \\
 & 3 & Visit Settings & Close & Allow & Opt-In \\
 & 3 & Visit Settings & Close & Do Not Allow & Opt-In \\
 & 3 & Visit Settings & Confirm my choice & Allow & Opt-In \\
 & \textbf{3} & \textbf{Visit Settings} & \textbf{Confirm my choice} & \textbf{Do Not Allow} & \textbf{Opt-Out} \\ \hline
\multirow{3}{*}{Opt-In} & 4 & How We Use Cookies & Close & NA & Opt-In \\
 & \textbf{4} & \textbf{How We Use Cookies} & \textbf{\begin{tabular}[c]{@{}c@{}}Opt-out of \\ google analytics\end{tabular}} & \textbf{NA} & \textbf{Opt-Out} \\
 & 4 & Thanks for letting me know & NA & NA & Opt-In \\ \hline
\multirow{5}{*}{Opt-In} & 5 & Accept All \& Close & NA & NA & Opt-In \\
 & 5 & Customize & Accept All & Allow & Opt-In \\
 & 5 & Customize & Accept All & Do Not Allow & Opt-In \\
 & 5 & Customize & Confirm my choice & Allow & Opt-In \\
 & \textbf{5} & \textbf{Customize} & \textbf{Confirm my choice} & \textbf{Do Not Allow} & \textbf{Opt-Out} \\ \hline
\multirow{4}{*}{Opt-In} & 6 & Do Not Sell My Data & Allow All \& Close & Allow & Opt-In \\
 & 6 & Do Not Sell My Data & Allow All \& Close & Do Not Allow & Opt-In \\
 & 6 & Do Not Sell My Data & Confirm my choice & Allow & Opt-In \\
 & \textbf{6} & \textbf{Do Not Sell My Data} & \textbf{Confirm my choice} & \textbf{Do Not Allow} & \textbf{Opt-Out} \\ \hline
\multirow{2}{*}{Opt-Out} & 7 & I'm in & NA & NA & Opt-In \\
 & \textbf{7} & \textbf{No, thank you!} & \textbf{NA} & \textbf{NA} & \textbf{Opt-Out} \\ \hline
\end{tabular}
\end{table*}

\subsubsection{Response Collection and interpretation}

Prior to commencing our study, we conducted an analysis of the cookie consent banners displayed on 20 different websites. We observed that online businesses intentionally design complex consent mechanisms, which hinders users' ability to make informed choices. The presence of multiple buttons and switches often confuses or frustrates users, leading them to hastily dismiss the consent banner without fully understanding its implications in terms of data sale. In our experiment website, we have recreated this environment to capture participants' authentic behavior.

For instance, in consent mechanism 1, clicking on the "Do Not Sell My Data" button does not automatically indicate that participants have opted out. Instead, they are presented with a second banner where they need to explicitly slide the toggle switch until it turns grey, and then click on "Confirm my choice" to record their response as "Opted-Out." If participants either click on "Close" or click on "Confirm my choice" without disabling the toggle switch in the second banner, their response will still be recorded as "Opted-In," even though they initially selected "Do Not Sell My Data." This demonstrates that the combination of buttons and switches in the subsequent step either impairs participants' ability to make an informed decision or frustrates them, leading them to hastily dismiss or ignore the banner.

Participants who chose to ignore the consent banner in the first or second step for consent mechanisms 1 through 6 were categorized as "Opted-In." According to the CCPA, businesses are permitted to sell consumers' data unless consumers explicitly withdraw their consent. Conversely, participants who disregarded the consent banner for consent mechanism 7 were classified as "Opted-Out," as Opt-In models do not assume consent by default. Participant responses were recorded in our experiment database, indexed with the unique Participant ID assigned to them in the presurvey questionnaire.

\subsection{Exit Survey}
The exit questionnaire consisted of Likert scale, multiple-choice questions and few open ended questions. It was divided into three sections -
\begin{enumerate}

\item \textbf{Reflection on Completed Activity} 
In this section, we prompted participants to recall and reflect on the online experiment. Firstly, we inquired whether participants had noticed the presence of the consent banner and whether they believed their behavior was being tracked for the purpose of selling data to third parties. Secondly, we asked participants whether they were provided with the option to Opt-Out of the sale or share of their data. Thirdly, participants were asked to rate their comfort level regarding the website's tracking of their behavior and the potential sale of their information to third parties. Lastly, we inquired whether participants could recall their choices made within the cookie consent banner and to explain their choices. 

\item \textbf{CCPA case examples}
Participants were presented with two hypothetical scenarios, which were constructed based on real privacy complaints investigated by the Office of the Attorney General at the State of California Department of Justice~\citep{ccpacase}. The first scenario revolved around registering for an online dating platform and explored whether clicking a share button when creating an account constituted sufficient consent for the sale of personal information, especially in cases where no additional "Do Not Sell My Personal Information" links were provided on the platform's homepage. The second scenario involved an online pet adoption platform where submitting an Opt-Out request necessitated a third-party authorized agent to submit a notarized verification on behalf of the user. These scenarios were accompanied by Likert-scale questions~\citep{274433} including: 1) "I think scenarios like this are likely to happen"; 2) "I would be concerned about my privacy in this scenario"; and 3) "Do you think the business acted appropriately and lawfully based on the situation described?" Additionally, participants were asked an open-ended question: 4) "Explain your reasoning above."
\item \textbf{Opt-In vs Opt-Out} 
Participants were initially queried regarding their familiarity with the distinction between Opt-In and Opt-Out consent mechanisms. Regardless of their prior response, they were subsequently provided with a debriefing explaining how each consent mechanism operates and its implications in terms of data sale. Participants were then asked to indicate their preferred consent mechanism. Following that, participants were given a debriefing on the economic implications \citep{Cate2001ProtectingPI} associated with Opt-In, and once again asked to specify their preferred consent mechanism. This economic implications debriefing can be found in the Opt-In vs Opt-Out section in the appendix.
\end{enumerate}
\vspace{-4mm}

\section{Data Analysis and Results}

In this section, we present participant demographics, compare the observations from our experiment with self-reported behavior in the exit survey, and discuss the results of our thematic analysis. Our findings provide insights into the level of concern or awareness among consumers regarding their privacy and privacy rights. Furthermore, we explore potential reasons why users are unable to successfully Opt-Out despite their intention to do so. To achieve this, we combine the findings from the online experiment, responses related to attitudes and concerns, reflections on the completed experiment, and explanations provided in a few open-ended questions. This comprehensive approach allows us to gain a deeper understanding of how users perceive the consent notices and why they make the choices they do. We reinforce our findings with participant quotes extracted from the responses to the open-ended questions.

After removing duplicate entries and rows with invalid data, a total of 360 participants responded to our invitation to participate in the study. 257 participants completed the primary task on our experiment website. After the primary task, we provided a link to an exit survey, which was completed by 232 participants. We utilized the data from the 257 participants who completed the primary task to examine the Opt-Out and Opt-In rates, as well as investigate the possible factors contributing to the low Opt-Out rates. 

\subsection{Participant Demographics}

Our participant pool consisted of 163 (63.4\%) males and 93 (36.2\%) females. The age range varied from 18 to 75 years old, with the most common age group being 25 to 34 years, which accounted for 166 (64.6\%) participants. The majority of participants, 106 (41.2\%), held a bachelor's degree as their highest level of education. While the highest number of participants, 92 (35.8\%), were from California, we also had participants from all other states in the US. The primary occupation for a significant portion of our participants was computer engineering, with 52 (20.2\%) individuals, suggesting a higher level of overall online literacy. Among the participants, the highest number, 126 (49.0\%), used a mobile device to complete the activity, while 167 (65.0\%) preferred Google Chrome as their browser, and 128 (49.8\%) used iOS/Mac as their operating system. In terms of self-reported internet usage, the most common range reported by participants was between 11 to 20 hours per week, with 94 (36.6\%) participants falling into this category.

\subsection{Experiment Results}
As a reminder, our experimental task primarily focused on participants indicating their preferred browser security indicators. Additionally, participants were randomly assigned one of the eight consent banners. They had the option to respond to the banners by clicking on the presented buttons or to ignore the banners altogether. Each participant's choices and the resulting interpretations were recorded and stored in our database.

\vspace{-4mm}
\subsubsection{Consent Mechanism Results}

Table~\ref{tab:countcm} provides a summary of the number of Opt-In and Opt-Out requests or preferences provided by participants on the experiment website. Our dataset is balanced due to the randomization and counterbalancing techniques discussed in the methodology, enabling us to compare the counts for each mechanism. It is important to note that our consent mechanism involved a two-step process.

Out of the total participants, we observed that 20.09\% (45/224) interacted with the first step of our consent mechanism. However, only 0.45\% (1/224) of users actually chose to Opt-Out in the second step for the consent mechanisms numbered 1 through 6. Consequently, 99.55\% (223/224) of participants did not Opt-Out, resulting in their consent decision remaining as Opt-In by default. This aligns with real-world websites governed by the CCPA, where users' data is considered to be sold unless they explicitly Opt-Out.

For consent mechanism 7, which is an Opt-In mechanism, we received only 4 Opt-In requests out of the 34 participants assigned to this mechanism. This indicates that only 11.76\% of users chose to Opt-In. Since this mechanism involved a one-step process, participants' first choice was sufficient to successfully Opt-In. In the next section, we will delve deeper into the analysis of our results.

\begin{table}[h]
	\scriptsize
    \centering
	\caption{Opt-Out and Opt-In counts for each consent mechanism. Here, \#Interact means the number of participant interact with cookie consent banner}
	\label{tab:countcm}
\begin{tabular}{@{}cccccccc@{}}
\toprule
\multicolumn{8}{c}{\textbf{Default Opt-In}}                 \\ \midrule 
\textbf{Mech.} &
  \textbf{\#N} &
  \textbf{\#Interact} &
  \textbf{\begin{tabular}[c]{@{}c@{}}1st Decision \\ Opt-In\end{tabular}} &
  \textbf{\begin{tabular}[c]{@{}c@{}}1st Decision \\ Opt-Out\end{tabular}} &
  \textbf{\begin{tabular}[c]{@{}c@{}}Default\\ Opt-In\end{tabular}} &
  \textbf{Opt-In} &
  \textbf{Opt-Out} \\ \midrule
1                     & 34  & 6  & 6  & 0  & 28  & 34  & 0  \\
2a                    & 32  & 8  & 6  & 2  & 24  & 32  & 0  \\
2b                    & 34  & 8  & 6  & 2  & 26  & 34  & 0  \\
3                     & 26  & 8  & 8  & 0  & 18  & 26  & 0  \\
4                     & 30  & 6  & 6  & 0  & 24  & 30  & 0  \\
5                     & 28  & 6  & 5  & 1  & 22  & 28  & 0  \\
6                     & 40  & 3  & 0  & 3  & 37  & 39  & 1  \\ \midrule
\textbf{Total Opt-In} & 224 & 45 & 37 & 8  & 179 & 223 & 1  \\ \midrule
\multicolumn{8}{c}{\textbf{Default Opt-Out}}                \\ \midrule
7                     & 33  & 6  & 4  & 2  & 27  & 4   & 29 \\ \midrule 
\textbf{Grand Total}  & 257 & 51 & 41 & 10 &  206   & 227 & 30 \\ \bottomrule
\end{tabular}%
\end{table}
\vspace{-4mm}

\subsubsection{Consent Mechanism Results Interpretation}
In the related work, we have highlighted the significance of even minor differences in implementation choices and their impact on how users perceive and respond to consent notices. Furthermore, existing literature emphasizes the importance of the question format. Although the data collected was evenly distributed among the eight mechanisms, we observed variations in the number of Opt-Out requests for each mechanism, as shown in Table \ref{tab:countcm}. In the following analysis, we will examine the influence of each consent mechanism on participants' consent decisions.

\textbf{2-Step Opt-Out Mechanisms} 
Consent mechanisms 1 through 6 involved a 2-step Opt-Out process. In the first step, if participants accepted, closed, or ignored the banner, it indicated that they had not opted out. In the second step, participants were asked to make a specific choice regarding the sale of their data if they clicked on other available options such as "Do Not Sell My Data," "Visit Settings," "Customize," or "How We Use Cookies". 

Table~\ref{tab:countcm} reveals that a total of 31 participants clicked on the buttons labeled "Accept Cookies \& Close,"(\#CM 2a,2b) "Close,"(\#CM 1) "OK,"(\#CM 3) or "Accept All \& Close"(\#CM 5) in the first step, indicating that they explicitly choose not to Opt-Out. This suggests that participants either made an informed decision, lacked sufficient understanding of online privacy, or were influenced by nudging factors that discouraged them from selecting the Opt-In buttons.

Furthermore, while 7 participants clicked on the more direct button "Do Not Sell My Data," one participant clicked on "Customize," and none on "How We Use Cookies" or "Visit Settings." This indicates that the direct buttons attracted more attention from users, while inline links were less commonly followed. In fact, this implementation choice, where businesses prioritize direct buttons over inline links, is one of the most common approaches employed by businesses (77.7\% of the top 5000 US websites in 2021) to discourage users from opting out more frequently \citep{DBLP:journals/corr/abs-2009-07884}.

As a result, out of the participants who clicked on the "Not Accepted" options (grouped as "Not Accepted") in the first step, only 1 participant proceeded to Opt-Out in the final step. Consequently, 97.78\% (44/45) of the participants who interacted with our Opt-In default consent banners remained in the Opt-In category. This suggests that although these participants did not immediately accept all cookies, they exhibited a higher level of curiosity or concern by exploring the other available options. However, their final decision did not align with their initial choice. Several factors could have influenced these decisions, including participants facing difficulty navigating the consent banners, altering their decision after reading the privacy policy, or losing interest in the privacy banner altogether.

\textbf{Nudging, Neutral and Anti-Nudging} 
Among the Opt-Out mechanisms developed in our study, all except for consent mechanism 2b and consent mechanism 6 employed nudging techniques. Consent mechanism 2b, a variation of 2a, utilized a neutral format, while consent mechanism 6 employed an anti-nudging approach. We hypothesized that the use of nudging could potentially manipulate users into selecting the highlighted option. In the case of consent mechanism 6, we expected that the highlighting of the "Do Not Sell My Data" button would lead to a higher Opt-Out rate. Similarly, we anticipated a relatively higher Opt-Out rate for consent mechanism 2b since both Opt-In and Opt-Out options were presented in the same format without any push towards a specific choice. However, in our study, we did not receive any Opt-Out requests for consent mechanism 2b. To substantiate this observation, a larger dataset would be required. The counts for each mechanism can be found in Table~\ref{tab:countcm}, presented above.

\textbf{Third-party Opt-Out Mechanism}
Consent mechanism 4 featured two inline links in the second step of the Opt-Out process. Six participants clicked on "Thanks for letting me know," indicating that they did not choose to explore the available options for managing their privacy preferences further. Only one participant clicked on "How We Use Cookies" in the first step and also on "opt out of Google Analytics tracking" in the second step. In our experiment, selecting "opt out of Google Analytics tracking" was interpreted as the user's intention to Opt-Out of the sale of their data. However, in real websites, clicking on a similar link would redirect users to a new page with instructions to download and install an add-on for their browsers. It was not possible to determine in this study whether the participant who clicked on "opt out of Google Analytics tracking" would actually proceed with the installation of the add-on.

\textbf{1-Step Opt-In Mechanism}
In consent mechanism 7, we introduced a default Opt-In consent banner with a 1-step Opt-In mechanism. We observed that 27 participants ignored the banner, and 2 participants clicked on "No, thank you!" This indicates that only a few users actively chose to Opt-In, suggesting that they are not readily willing to share their data with businesses when they are not assumed to be Opt-In by default. However, under the current default model of CCPA, users' data remains accessible because the process of opting out can be confusing or cumbersome. On the other hand, 4 participants clicked on "I'm in." On real websites, clicking on similar buttons would provide businesses with explicit consent to sell or use their data. Although the number of participants who provided express consent in our experiment is small, it is noteworthy that these participants granted consent to an unknown website they visited for a research study on browser indicators. This could be attributed to participants' lack of online privacy literacy or the influence of nudging techniques.

We analyzed participants' explanations for their consent decisions (Section: Reflection on Completed Activity) and quote few of them that represent the most commonly reported reasons for:
\vspace{-2mm}
\begin{enumerate}
\item Accepting Cookies - "If you do not select Accept, the site will not function properly", "Automatic click to the big button that says accept.","Cause this one takes less time" and "I'm open to resource sharing".
\item Rejecting Cookies - "This is my personal data so I don't agree to sell it", "I pay more attention to information security", "I'm afraid they're selling it like crazy" and "I don't want to reveal my privacy to the outside world. I don't feel good about it".
\end{enumerate}
\vspace{-6mm}
\subsubsection{Statistical Analysis}
Significance tests were conducted for the below between-group studies. We used consent decision as the dependent variable and the miscellaneous factors as the independent variables and a 95\% confidence interval for all our significance tests.
\vspace{-4mm}
\paragraph{Residential Status in California}
Since CCPA applies only to residents and businesses in California and only two other states in the US have similar privacy protection laws, we compare user behaviors from different states. 11.96\% participants from California and 11.52\% participants from all other states (49 states, Puerto Rico and District of Columbia) opted-out. 
A one-way ANOVA test suggests that there is no significant difference among how users from different states and territories in the US respond to consent notices (F(39,217) = 0.792, \textit{p} = .805). However, as we discuss later in our survey results, 74.70\% participants from California reported they are slightly to extremely concerned about their privacy and are not comfortable sharing their data with businesses.

\vspace{-4mm}
\subsubsection{Miscellaneous platform factors} We have learned that the design or implementation of the consent notices, or in other words how the consent banners are displayed on users' screens, impact users' choices or their ability to make these choices. The consent banners may have slight variations in look and feel depending upon the device, browser or operating system used. Using a one-way Anova test, we found that there was no observed statistically significant difference between groups of devices (F(3,253) = 0.461, \textit{p} = .710), browsers (F(5,251) = 0.962, \textit{p} = .442) and operating systems (F(3,253) = 0.574, \textit{p} = .632).

\vspace{-4mm}
\subsubsection{Miscellaneous demographic factors}
We analyze the impact of demographic factors into the consent decisions. We found that there was no statistically significant difference in the number of Opt-Outs or Opt-Ins when comparing males and females (t=0.847, df=254 \textit{p} = 0.389), level of education (F(7,249) = 0.597, \textit{p} = .759), hours spent on the internet (F(4,252) = 1.439, \textit{p} = .221), occupation (F(13,243) = 0.844, \textit{p} = .613), and age groups (F(4,252) = 0.413, \textit{p} = .799).

\subsection{Exit Survey Result}
In the following section we present the results of the exit survey for Reflection on completed activity, CCPA case examples, and Opt-In vs Opt-Out.

\vspace{-4mm}
\subsubsection{Reflection on Completed Activity}

In this section, we will analyze the participants' reflections on their completed activity. A majority of the participants, 71\%, were able to recall that our experiment website provided them with the option to Opt-Out of the sale of their personal data. When asked about how often they notice websites offering the option to opt-out of data sale, 32.5\% of participants stated that they rarely or never notice this option. Regarding their choice in the experiment, 57.2\% of participants mentioned that they either accepted the cookies or closed the Opt-In consent banner. Additionally, 31.6\% of participants indicated that they chose not to sell their data. The remaining participants were unsure about their choices or mentioned visiting the settings. In terms of comfort level, 14.7\% of participants stated that they would be very comfortable if the experiment website tracked their behavior and sold their information to third parties, while 16.9\% expressed being very uncomfortable with this idea.

\begin{figure}[ht]
    \centering
    \includegraphics[width=\linewidth]{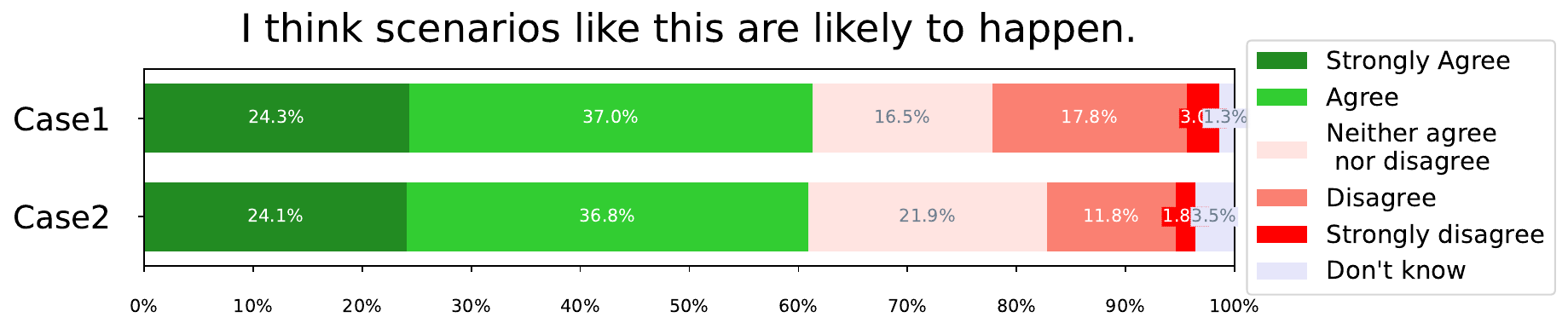}
    \includegraphics[width=\linewidth]{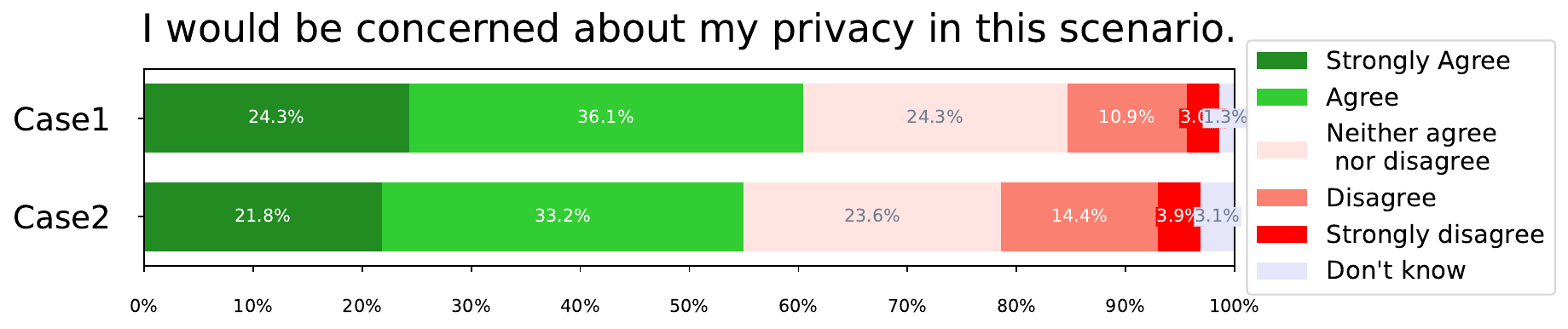}
    \includegraphics[clip, trim=0cm 0cm -2.6cm 0cm,width=\linewidth]{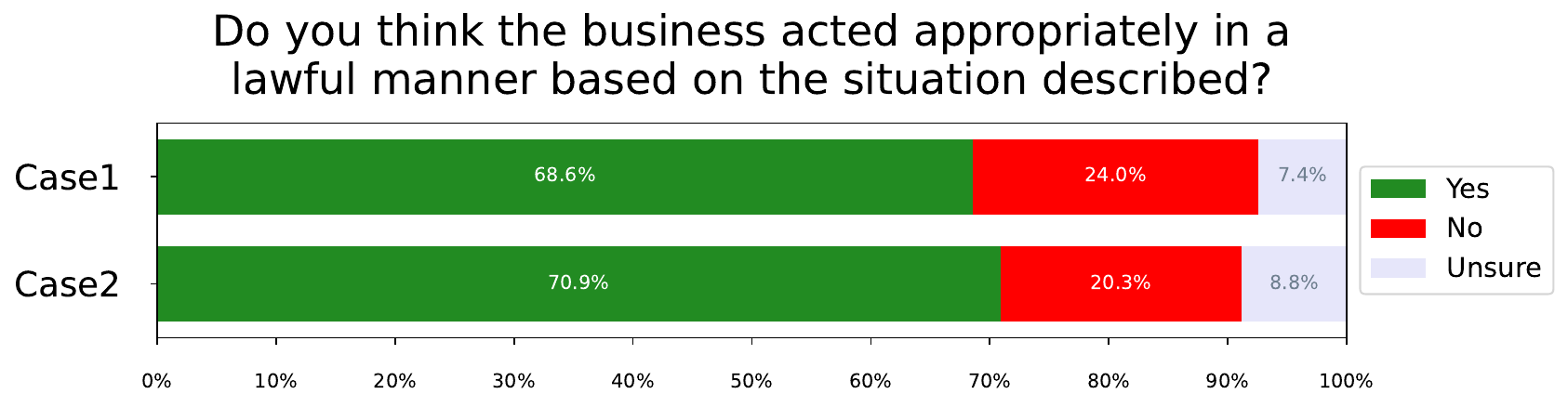}
    \caption{Participants' responses to CCPA case examples.} 
    \label{tab:c6}
\end{figure}

\subsubsection{CCPA case examples}

The cases presented in this study depict two distinct scenarios involving how businesses handle and sell customer personal information, as well as a user's ability to Opt-Out of the sale or sharing of their data. While a majority of participants expressed concern in response to these scenarios, only 24\% and 20.3\% of participants correctly identified scenarios 1 and 2 as unlawful (see Fig~\ref{tab:c6}). Many participants felt that since users were "warned in advance," the practices were deemed acceptable. This finding suggests that businesses can potentially exploit consumers' limited awareness to their advantage. It is important to note that these scenarios were based on real CCPA cases, in which the Office of the Attorney General notified the companies of alleged noncompliance and corrective actions were subsequently taken~\citep{ccpacase}. This observation highlights the lack of knowledge among users regarding their privacy rights.

\subsubsection{Opt-In vs Opt-Out}

A significant majority of participants, 85.7\%, reported that they were aware of the distinction between Opt-In and Opt-Out models. After being presented with the differences between these models, 82.0\% of participants indicated a preference for Opt-In. We found that 3.1\% of participants changed their preference to Opt-Out after being informed about the potential economic implications associated with the Opt-In model. This suggests that 78.9\% of participants still preferred Opt-In despite the awareness of these implications. Although there was a slight decrease in the number of individuals favoring Opt-In after being briefed about the economic implications, the overall preference for Opt-In remained higher. This could be attributed to users' high level of concern for their privacy and their inclination towards Opt-In in all circumstances. Another possible explanation is that participants gained valuable insights into the use of their data by businesses, as well as the Opt-Out and Opt-In systems, throughout the study, leading to heightened concerns about their privacy.

\section{Discussion}
\subsection{Understanding User Consent Choices and Privacy Actions}

Out of the 224 participants assigned to the Opt-Out consent mechanisms (1-6), 179 participants chose to ignore the consent banner. Additionally, 8 participants did not Opt-Out despite not accepting the cookies in the first step, and only 1 participant actually opted out in the subsequent step. 

When we asked participants to explain their reasoning, the most common response was that they clicked on the "Do not sell my data" button because of concerns about privacy and their personal data. However, not all users who click on this button successfully Opt-Out. 

Furthermore, we observed that the most commonly used implementation choices have a negative impact on Opt-Out rates, either due to the difficulty of navigating through multiple options or the influence of nudging techniques.

Although our study did not include consent mechanisms that require users to fill out forms or send requests via email, it is likely that even fewer users would go through the process due to the time-consuming nature of these methods. For example, one of our participants mentioned that they clicked on "Accept cookies \& Close" because it was a quick option. Businesses have the ability to reduce Opt-Out rates through various design choices, as only a small percentage of users have the patience or willingness to complete the entire Opt-Out process.

Furthermore, we observed that very few users file complaints, and even fewer are able to identify if a website is acting unlawfully. Without significant pressure from users, companies and policymakers are likely to maintain the status quo and neglect necessary corrective changes.

\subsection{Exploring User Interaction with Consent Banners}

The Opt-Out mechanism, which assumes consent by default, typically has lower Opt-Out rates, and users remain opted-in. However, this does not necessarily mean that all users made an informed decision, fully understood the implications, or were aware of how the process works if they did not click on any button. When given an explicit option to Opt-In, users seldom choose to do so. This suggests that in an Opt-In regime, only a small number of users would actively opt-in for the use of their data, due to the default nature of the Opt-Out system.

Users generally prefer the Opt-In model because it provides them with power and control over their data. Interestingly, users rarely interacted with the consent banners at all. This aligns with the findings of O'Connor et al.~\citep{DBLP:journals/corr/abs-2009-07884}, who observed a maximum interaction rate of 20.5\% with their consent banners, compared with 19.84\% in ours. In both Opt-In and Opt-Out scenarios, inaction was the most common outcome. By comparing the outcomes of Opt-In and Opt-Out banners, we can see that they lead to significantly different results based on the default condition. While further research is needed to explore the specific reasons for this inaction, it is clear that users' preferred mechanism differs from their chosen one due to the complexities of websites.

\subsection{Limitations and Future Work}
Given the small size of our sample population, the statistical power of our results is limited. To ensure the validity of these findings, it is necessary to replicate this study with a larger and more diverse sample that is representative of the general population. It is important to note that our survey results may not fully capture the tendencies of US residents outside of California, as CCPA primarily applies to California. Furthermore, the nature of our study being conducted within a university setting and utilizing a survey format may introduce biases, attracting younger and more educated participants who may be more inclined to consent to sharing their information compared to a real-world advertising context. Additionally, since our study was conducted exclusively in the US, the findings may not be applicable to other regions around the world. Furthermore, its important that we acknowledge the effect that utilizing pure HTML in our website design might have on the consent choices of participants. These choices contributed to an outdated appearance of our pages, and may have potentially influenced participant interaction rates with the banners or choices to consent.
Future studies should aim to replicate these experiments with a non-Western, Educated, Industrialized, Rich, and Democratic (non-WEIRD) population to explore their perspectives and investigate potential differences between WEIRD and non-WEIRD populations regarding data privacy. 
Specifically, future research can tailor scenarios where participants have no pre-existing trust relationship with the website and where security indicators are not the primary focus. This would allow for a more comprehensive understanding of user behavior and decision-making in relation to consent banners.

\subsection{Recommendations to Policy Makers}

The feasibility of adopting the Opt-In model in the US market should be further explored. Under the current Opt-Out model, most users remain opted-in not because they want to, but due to either a lack of knowledge or the tedious and confusing Opt-Out process. The Opt-In model, as seen in GDPR, offers several benefits. Businesses can maintain a lean database of highly relevant leads who are genuinely interested, reducing data management overhead. Brands that are transparent and prioritize their customers' privacy gain consumer trust and can build stronger relationships. Adopting an Opt-In model can foster trust, open new opportunities, and provide internet users with a greater sense of safety and control over their data, without having to go through additional hurdles.

While the Opt-Out process under CCPA may seem promising, our study reveals that it is still far from effectively addressing consumer preferences. Standardizing the Opt-Out process is crucial. By establishing consistent formats, users will be relieved from the burden of navigating complex privacy forms and successfully opting out. Additionally, within this standardized setup, there should always be a single-step option available for users to easily opt out. Strict monitoring of businesses' compliance is necessary to ensure that users can make informed decisions based on their preferences, without the need to navigate through multiple buttons or web pages to submit an Opt-Out request. Awareness campaigns should also be launched to educate consumers about privacy laws, complaint filing procedures, and their privacy rights through mediums like radio and television. It is imperative that privacy laws, whether Opt-Out or Opt-In, are introduced in all states across the US to ensure consistency and protection for all consumers.
\section{Conclusion}

In this study, we conducted a deceptive experiment to evaluate user responses to CCPA compliant cookie consent banners. Our findings indicate that only 0.45\% of participants chose to Opt-Out in the default opt-in model. Despite expressing a desire to Opt-Out, the current implementation of the Opt-Out mechanism hinders users from successfully doing so. Conversely, in the default Opt-Out mechanism, only 12.12\% of participants opted-in. These results reveal a discrepancy between users' self-reported preferences to Opt-Out and the actual outcomes observed in our study. To address this issue, policymakers should establish clear guidelines for companies to follow in implementing the Opt-Out or Opt-In mechanisms, ensuring a standardized approach rather than allowing for variations in the steps involved.

\section{Acknowledgement}
ChatGPT was utilized to rectify grammatical errors and enhance the clarity and readability of the entire text. Primarily, we have used a common prompt: "Please correct grammatical errors, and improve the readability and clarity of this paragraph." We extend our gratitude to all study participants for their time. Furthermore, we are grateful to the anonymous reviewers for improving our paper for publication.

{\small{
\bibliographystyle{splncs04}
\bibliography{main}
}}
\onecolumn
\section*{APPENDIX A}

\textbf{Reflection on Completed Activity}

\begin{enumerate}
   \item Answer the following (Yes, No, Unsure)
   \begin{enumerate}
    \item Did the website you visited for this activity track your behavior and sell this information to third parties?
    \item Did the website you visited for this activity give you an option to opt out of the sale of your personal data?
   \end{enumerate}
   \item If this website tracked your behavior and sold this information to third parties, how comfortable would you be with it? (Very Comfortable, Somewhat comfortable, Neutral, Somewhat uncomfortable, Very uncomfortable)
   \item Which option for consent did you choose? (Do not sell my data, Opt-In, Close, Accept Cookies and Close, OK, Visit Settings, Unsure)
 \end{enumerate}

\noindent\textbf{CCPA case examples} 

Imagine yourself in each of the following scenarios and indicate to what extent you agree or disagree with each statement.
\begin{enumerate}
    \item You have registered on an online dating platform. A user clicking an “accept sharing” button when creating a new account is sufficient to establish blanket consent to sell personal information as per this business. There is no additional “Do Not Sell My Personal Information” link on its homepage.
    \begin{enumerate}
        \item I think scenarios like this are likely to happen. (Strongly Agree, Agree, Neither agree nor disagree, Disagree, Strongly Disagree, Don’t know)
        \item I would be concerned about my privacy in this scenario. (Strongly Agree, Agree, Neither agree nor disagree, Disagree, Strongly Disagree, Don’t know)
        \item Do you think the business acted appropriately in a lawful manner based on the situation described? ( Yes, No, Unsure)
        \item Explain your reasoning above. [Textbox]
    \end{enumerate}
    \item A business that operates an online pet adoption platform requires your authorized agent to submit a notarized verification when invoking your privacy rights. The business directs you to a third-party trade association’s tool in order to submit an opt-out request.
    \begin{enumerate}
        \item I think scenarios like this are likely to happen. (Strongly Agree, Agree, Neither agree nor disagree, Disagree, Strongly Disagree, Don’t know)
        \item I would be concerned about my privacy in this scenario. (Strongly Agree, Agree, Neither agree nor disagree, Disagree, Strongly Disagree, Don’t know)
        \item Do you think the business acted appropriately in a lawful manner based on the situation described? ( Yes, No, Unsure)
        \item Explain your reasoning above. [Textbox]        
    \end{enumerate}
\end{enumerate}

\noindent\textbf{Opt-In vs Opt-Out} 
\label{opt-in-opt-out-app}
\begin{enumerate}
    \item Do you understand the difference between Opt-in and Opt-out? (Yes, No, Unsure)
    \item Which of the below options would you rather have businesses follow?
    \begin{itemize}
        \item Option A : “Opt-In”
    \end{itemize}
    \begin{enumerate}
        \item Default settings: Do not sell data
        \item Explicitly ask user for consent before selling data
        \item If user doesn’t provide consent, do not sell data
        \item User will not get customized recommendations
        \item User will not get directed advertisements
    \end{enumerate}
    \begin{itemize}
        \item Option B : “Opt-Out”
    \end{itemize}
    \begin{enumerate}
        \item Default settings: Sell data
        \item Ask user if they want to revoke consent
        \item If user revokes consent, do not sell data
        \item By default, user will get the full experience of the service, customized recommendations and get directed advertisements
    \end{enumerate}
    \item Research says "opt-in" impedes economic growth by raising the costs of
providing services and consequently decreasing the range of products and services
available to consumers. It would make it more difficult for new and often more
innovative, firms and organizations to enter markets and compete. It would also make it
more difficult for companies to authenticate customers and verify account balances. As a  result, prices for many products would likely rise.
Which option would you prefer with the information presented above? (Opt-In, Opt-Out)
\end{enumerate}

\end{document}